\documentclass[article,twocolumn]{revtex4}

\usepackage{amssymb,amsmath,amsfonts}
\usepackage{graphicx}

\begin{document}

\title{An Integrated Approach to Understanding Vacuum Arcs}

\author{J. Norem$^*$}
\affiliation{Nano Synergy Inc., Downers Grove, IL, USA}
 \email{norem.jim@gmail.com}
 
\author{Z.  Insepov}
\affiliation {National Research Nuclear University (MEPhI), Kashira Hwy, 31, Moscow, Russia, 115409}
\affiliation{Purdue University, West Lafayette, IN, USA}

 \author{A. Hassanein}
\affiliation{Purdue University, West Lafayette IN, USA}
 
\date{\today}
 
\begin{abstract}
Although used in the design and costing of large projects such as linear colliders and tokamaks, the theory of vacuum arcs and gradient limits is not well understood.   Almost 120 years after the isolation of vacuum arcs, the exact mechanisms of the arcs and the damage they produce still being debated.  We describe our simple and general model of the vacuum arc that can incorporate all active mechanisms and aims to explain all relevant data. Our four stage model, is based on experiments done at 805 MHz with a variety of cavity geometries, magnetic fields, and experimental techniques as well as data from Atom Probe Tomography and failure analysis of microelectronics.  The model considers the trigger, plasma formation, plasma evolution and surface damage phases of the arc.   Our data clearly shows surface damage produced by differential cooling capable of producing local high field enhancements $\beta \sim 200$, and arcing in subsequent pulses.   We update the model and discuss new features while also pointing out where new data would be useful in extending the model to a wider range of frequencies.
\end{abstract}

\maketitle

\section{Introduction}                       

Arcing on surfaces occurs in many environments under many different initial conditions, such as DC, RF, vacuum or gas, pre-existing plasma, wide or narrow gap, clean or dirty conductors 
\cite{C,E,A,K,B,H,alpert}, for normal and superconducting systems \cite{J}, with and without strong B fields \cite{PR1,PR2}, and ignited by high electric fields, lasers \cite{laser} or particle collisions \cite{astro}.  Although vacuum arcs have been studied for over 120 years \cite{earhart,Hobbs}, and the first credible model of single surface electrical breakdown was Lord Kelvin's argument in 1904 that local fields on the order of 10 GV/m would produce mechanical failure  \cite{lordk}.  This prediction assumed that local surface fields could be many times higher than the average surface field, now called the field enhancement factor, $\beta$, described by Alpert, accurately estimated the local field at breakdown for a range of data \cite{alpert}.  In the "Feynman Lectures" a simple derivation shows that, while high $\beta$ values can be created by "fencepost" geometries, surface fields may be primarily determined by the local curvature of the surface rather than the larger features of asperities \cite{feynman}.  

In 2001, J\"{u}ttner summarized the theoretical understanding of vacuum arcs in a review, where he argued that the understanding of arcing had not converged on a single theory applicable to a wide variety of applications, and much of the active effort in the field produced contradictory conclusions and disagreement \cite{Juttner}.  Our argument, \cite{IN}, is that a simple model may be the basis for a useful understanding of the process.  

The most commonly used breakdown model is the Explosive Electron Emission (EEE) system based on studies of field emission heating of asperities \cite{C,dyke,A} although wire shaped asperities are not seen and the rf duty cycle for field emission is only about 0.1 \cite{PR1}.  Meanwhile, other mechanisms that could trigger breakdown events are seen in a variety of environments.  For example, the EEE model applies only to cathodic arcs, however similar surface fields at positive potentials (with no field emission or heating) also mechanically fracture surfaces at roughly the same local fields \cite{F,G} in Atom Probe systems.

Our primary interest has been the study of mechanisms limiting the accelerating gradients of modern accelerators, since the overall cost of linear accelerator facilities is related to the gradient that can be maintained in these structures, primarily because lower gradients mean longer structures are required to produce the required performance \cite{RF}.  In the design of tokamak power reactors, arcing can introduce impurities into the plasma to compromise the operation of the systems, as well as limiting the power that is available to heat the plasma to the temperature required for fusion \cite{ITER}.  The design of high voltage transmission lines is limited by the constraint that high surface fields can produce corona discharges that limit the operational voltage and ultimately are directly responsible for the loss of approximately 4\% of the transmitted power, with the associated costs and pollution \cite{EPRI,cor,HVTL}.  Although less obvious, we find that the physics limiting integrated circuit design may also be related to surface arcing since high current densities seem to be the trigger both in electronic component failure \cite{D,DD} and vacuum arcs.

Surface arcs are difficult to study, and many efforts seemed to be interested in optimizing performance of systems rather than basic plasma physics.  Arcing can operate extremely rapidly (ns scale), the dimensions involved may be very small (a few microns or less),  the dynamic range of many parameters may extend over many orders of magnitude, and the scale of surface damage from arcs and surface asperities leading to arcing may be on the order of a few nm.  A further problem is the environment of experimental measurements, which must involve factors such as gas, vacuum, high fields, and unpredictable arc positions.  Further complicating understanding this physics is the problem that the results of arcing experiments seem to be nonlinear due to thresholds in arc duration, available energy and other variables, so that measurable parameters seen in one experiment may not be detectable in experiments done with somewhat different parameters.  

Although the conventional wisdom seems to be that "{\it theoretical understanding of breakdown remains incomplete at this time, particularly with regards to the micro- structural mechanisms of the field-induced ejection of matter from the structure surface that initiates the evolution of breakdown plasma, and with regards to how the mechanisms are affected by structure material conditions} \cite{dj}", this paper argues that a realistic model of breakdown can be simple, accessible and useful.

\section{Breakdown without Heating}    

Since "fencepost" (unicorn horn) shaped asperities are not seen experimentally, we argue that a more general model of breakdown is needed using mechanisms where Ohmic heating is not required, as shown in Fig. 1.   We assume that breakdown occurs when Maxwell stresses are greater than the tensile strength of the material, which occurs at surface cracks or small local radii \cite{IN}.   These processes occur in a wide variety of well studied environments.

\begin{figure}[htb]  
   \centering
   \includegraphics[width=8cm]{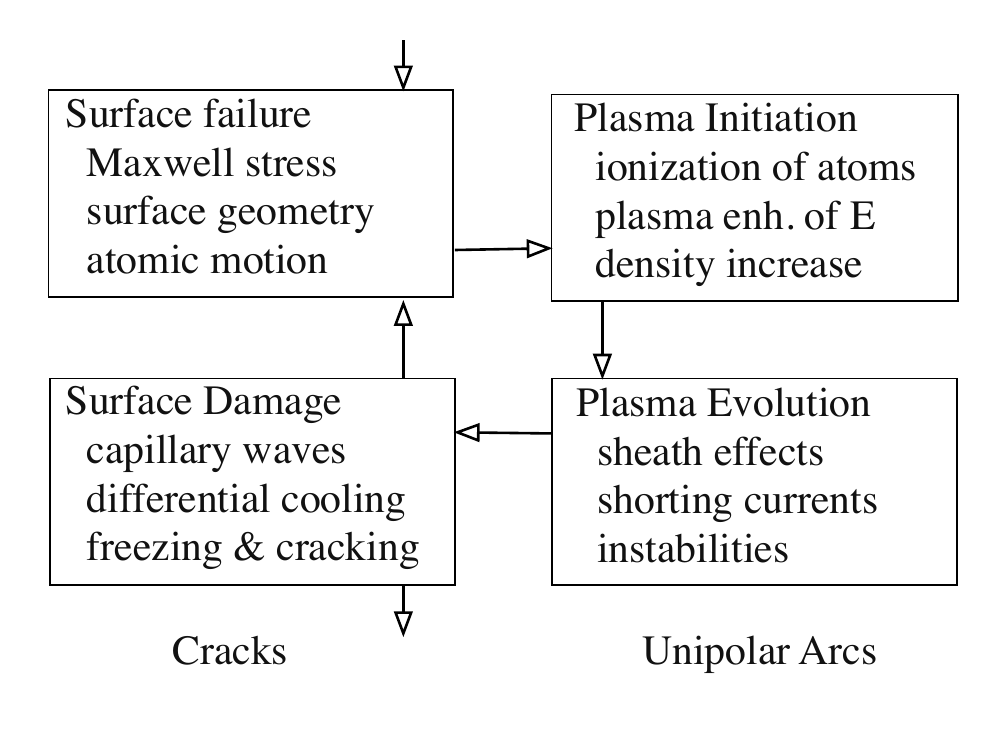}
   \caption{Vacuum arc development involves 4 stages.  We consider processes that seem dominant at different stages of the development of the arc, and find that under continued operation the arc follows a life-cycle, where damage from one breakdown event is very likely to produce another.  We also find that cracks due to differential cooling and unipolar arc physics explain much of the experimental data we see.}
\end{figure}

In this model the lifecycle of the arc as divided into four parts: trigger, plasma ionization, plasma evolution, and surface damage.  The process occurs in four stages: 1) local surface fields, measured from field emission, are high enough so that Maxwell stresses can be comparable to tensile strength causing surface failure \cite{lordk,PR1}, 2) field emission ionizes the fragments of surface material, producing a positively charged ion cloud near a field emitter that will increase the field on the emitter \cite{Noremrf2011,Schwirzke91,Robson}, 3) an unstable. non-Debye plasma is maintained by field emission and self-sputtering \cite{morozov,insepovsput,noremIPAC12,a92}, and, 4) surface damage is caused by Maxwell stresses, thermal gradients, and surface tension on the liquid metal surface.\cite{IN,LL,hassanein}

Our model is primarily based on data taken at 805 MHz with and without 3 T co-linear B fields, done as part of the Fermilab contribution to the Muon Accelerator Project, during 2001 - 2012 \cite{PR1,IN}.  All numerical results in this paper assume a copper structure.

\subsection{Triggers}          

We assume that high Maxwell tensile stresses mechanically break the surface, producing a local cloud of neutral atoms close to the asperity, which continues to field emit.  The parameters of breakdown events obtained in breakdown studies at 805 MHz, are shown in Fig. 2 \cite{PR1}.  

\begin{figure}[htb]  
   \centering
   \includegraphics[width=8cm]{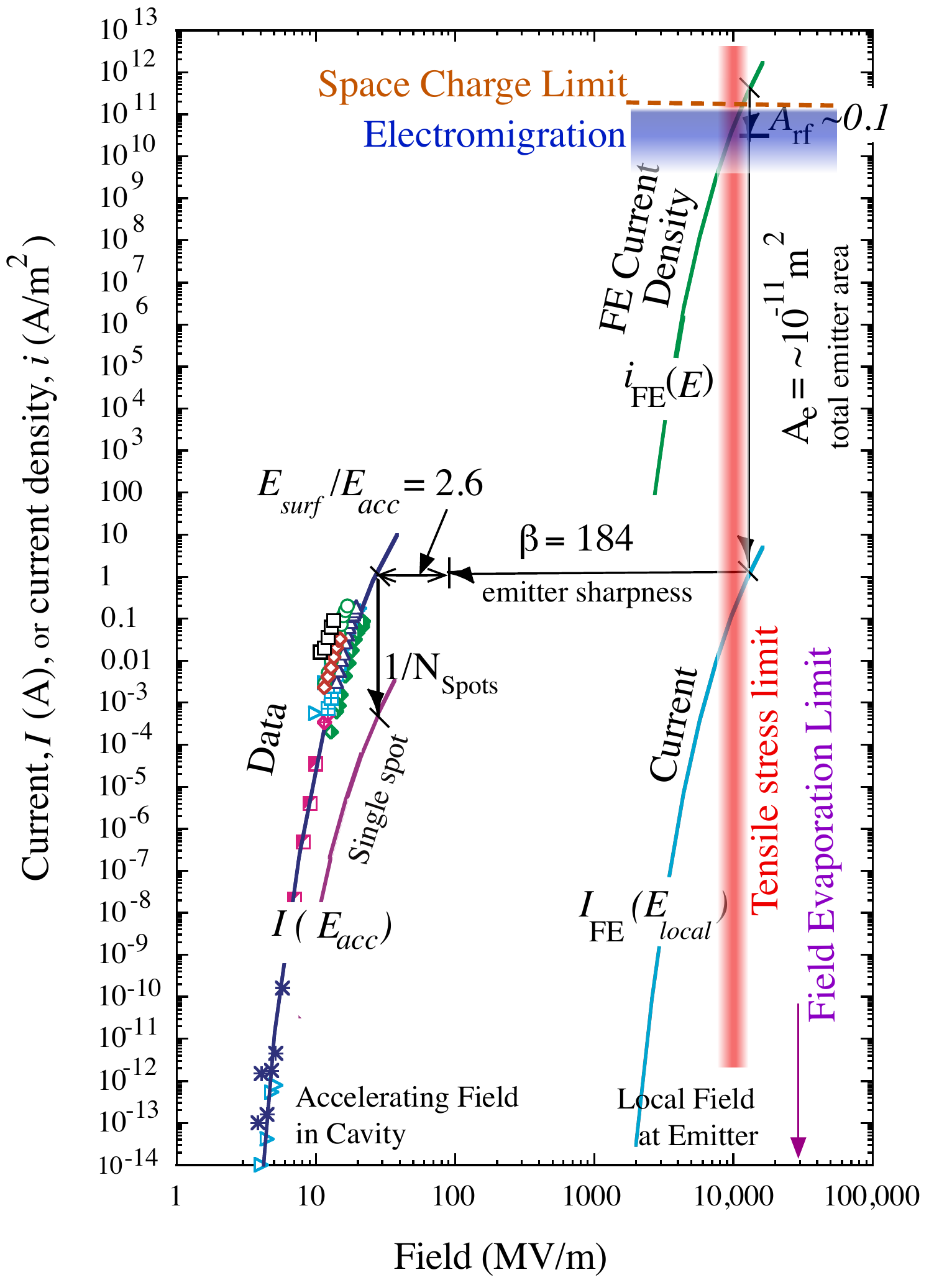}
   \caption{Experimental data of field emission current in a cavity \cite{PR1} compared with fitted models, showing the magnitude of current density \cite{H} and electric field at the field emission sites, along with the limits that would be imposed by the material tensile strength \cite{PR1}, electromigration \cite{D}, and field evaporation from smooth surfaces \cite{F} and the space charge limit \cite{dyke,IN}.  The experimental data sums the contributions of roughly 1000 emitters \cite{PR1}.  The precise values of the various limits depend on the metallic properties and experimental conditions.}
\end{figure}

Following Lord Kelvin, this model assumes that triggers are due to mechanical failure of the surface due to Maxwell stresses comparable to the tensile strength and the high fields required are produced the corners that exist at crack junctions and other features \cite{lordk,insepov,IN}.   In practice, the arguments of Lord Kelvin are difficult to apply at atomic dimensions, and it is necessary to use more sophisticated models and mechanisms.  Failures of metallic samples in Atom Probe Tomography Systems, where visible flashes are seen without the presence of field emitted electrons imply that field emission is not necessary to produce surface failure, see Section 5.7 of Ref. \cite{F}.  Thus triggers and particulate production could occur with a positively charged surface 

In rf experiments, the values of accelerating fields can be determined from the geometry and applied power.  The local field at the breakdown location is more difficult to determine but can be measured from the dependence of field emitted current on electric field, which depends on the exponent $n$,  $I_{FE} \sim E^n$ The values of electric field for cavities and local models are shown in Fig. 3 \cite{PR1}.  Local asperities enhance the local field on the surface by a factor 
\[\beta = E_{local}/E_{surf} \sim 50 - 1000,\] 
where the enhanced field on the asperity $E_{local}$, and the average surface field on the surrounding area is $E_{surf}$, following Ref. \cite{alpert}.

It is a useful  oversimplification to say that a surface will break down at E$_{local} \simeq 10$ GV/m. Because all real surfaces we consider are rough, possibly under stress at some level, and are not clean, the work functions $\phi$ are not precisely known and the spectrum of field enhancements due to multiple asperities are not well known.  We determine the local field on the emitter using the field emission model of Brody and Spindt \cite{F}, used in vacuum microelectronics.  We assume that the work function $\phi \simeq 6$ eV, because the primary impurity  present would be oxygen, which is electronegative \cite{H}.  This is shown in Fig 3.  The experimental parameters of the field emitting surface are limited because the only measurable quantity during high gradient operation is the exponential dependence of the field emitted current on field, requiring an estimate of the surface work function to determine the local field. 

It is not possible to assume that the surface geometry of asperities is constant, since high local fields put significant stress on surface atoms.  Many models assume that field enhanced diffusion explains how the field enhancement of asperities could increase with time, however electromigration (described below), which dominates diffusion under low field conditions, should maintain its relative dominance over diffusion under high stresses, as shown in experiments, see Table 5.23 of  \cite{D} and Tables 2.1 and 4.5 of \cite{DD}.   

\begin{figure}[htb]  
   \centering
   \includegraphics[width=8cm]{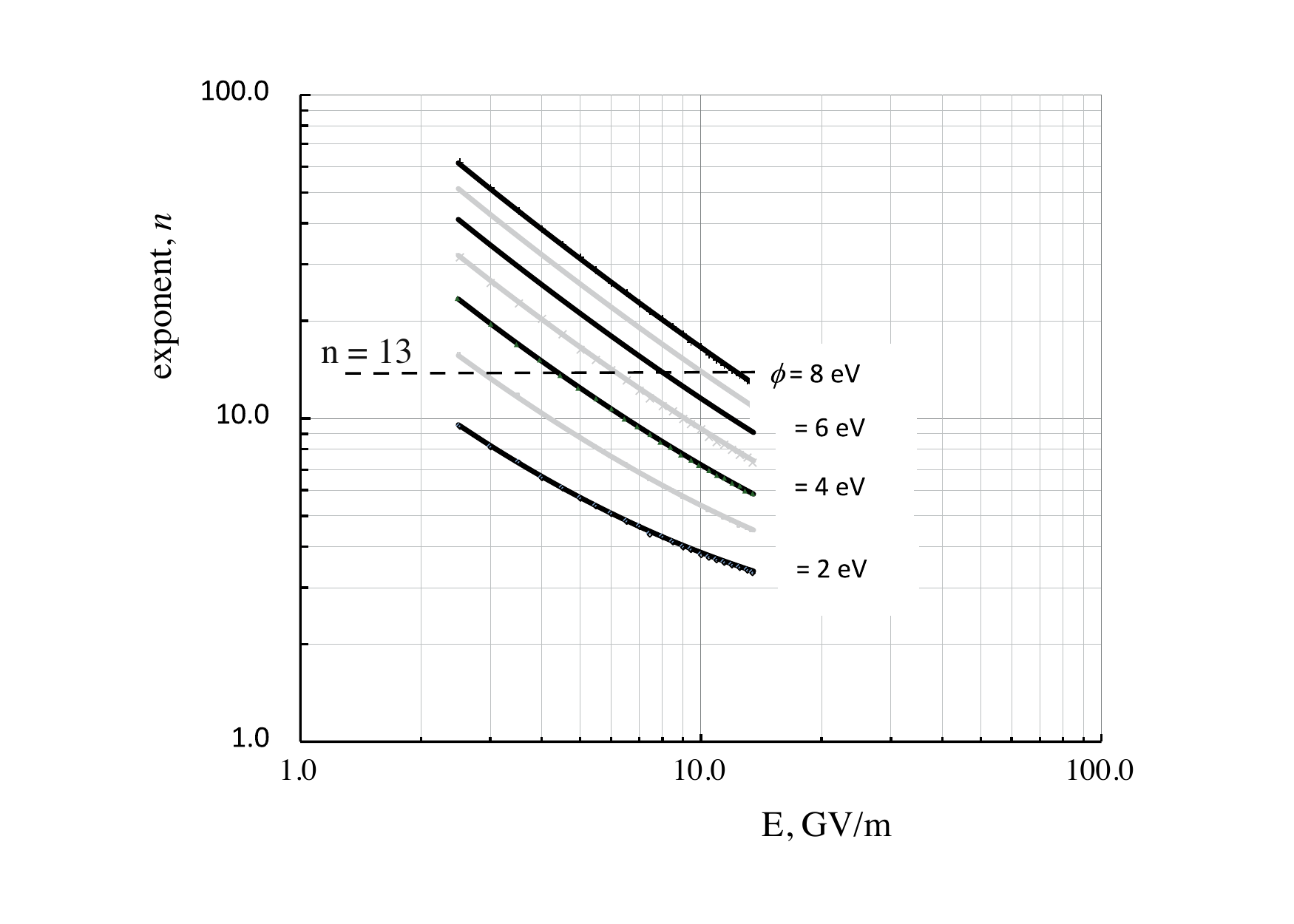}
   \caption{The local electric field, $E_{local}$, can be obtained from the exponent $n$ from field emission measurements $I \sim E^n$ and estimates of the work function $\phi$, \cite{F}.  Experimental data shows that $n \sim 14$ for radiation levels and $n\sim 13$ for field emission \cite{PR1}}.
\end{figure}

\subsection{Plasma Initiation}          

The material removed from the surface would continue to be exposed to field emission, which would ionize it.  This model assumes that plasma ions are produced with very low ion temperatures and essentially confined inertially, producing a low temperature plasma near the field emitter, trapping some of the electrons, but leaving the plasma with a net positive charge.  The resulting sheath potential (and image charge) increases the field on the field emitter and confines the plasma close to the surface.  Field emission maintains the plasma electrons \cite{Noremrf2011}.  Self-sputtering from plasma ions maintains the ion density \cite{insepovsput},  Image charges provide the charge to stick the plasma to the surface. And the density rises until the plasma plasma becomes non-linear (non-Debye) \cite{morozov}.

Particle in Cell, (PIC) codes show that the initial plasma temperature, both $T_e$ and $T_I$ is cool, only a few eV, however the sheath potential between the plasma and the walls can be significantly higher than the plasma temperature.  This relationship persists as the density increases and the system eventually becomes non-Debye \cite{morozov}.   Arc damage implies the arcs are round and roughly 500 $\mu$m in diameter. 

\subsection{Plasma Evolution}          

We assume that the plasma produced in arcs is a unipolar arc, first described by Robson and  Thoneman \cite{Robson}. in 1959, and later by Schwirzke \cite{Schwirzke91}, Anders \cite{C} and J\"{u}ttner  \cite{Juttner}.  The arcs are dense, unstable and frequently in motion.  There is an extensive literature on these arcs, the damage they produce and their complex behavior.   Although classical unipolar arcs have no external currents, field emission at the surface could produce a net current, and could be a source of instabilities.  

The properties of unipolar arcs depend primarily on the plasma density, which can be high enough to make the plasma non-Debye, which occurs at around $6 \times 10^{26}$ m$^{-3}$, see Fig 3.48 in ref \cite{C} and \cite{a92}.  The arc properties are strongly dependent on the plasma sheath \cite{morozov,noremIPAC12}, particularly ion self sputtering at high temperatures and high tensile stresses.  When combined with other existing data on arc behavior, modeling using PIC and Molecular Dynamics (MD) codes, see Fig 4, has shown that densities in this range can explain the gross features of the Debye lengths, burn voltages (sheath potentials) and other plasma properties measured experimentally.  

In Fig. 4, MD was used to evaluate the local equilibrium electron densities produced when electrons, which move much faster than ions, leave the plasma boundary, producing the sheath that depends primarily on the electron temperature and ion density.  Calculations of the sheath of non-Debye plasmas at high densities have shown that electron temperatures, $T_e \sim 3$ eV and ion temperatures less than this would be consistent with experimental data on plasma density and burn voltages of 23 V, seen experimentally \cite{IN,morozov} and Table B8 of Ref \cite{C}.   The calculations predict that the surface electric field produced in the sheath would be on the order of  $E > 7 \times 10^9$ V/m, sufficient to produce significant field emission from flat surfaces without any field enhancement. These currents would also be sufficient to short out the driving field and absorb all the electromagnetic energy in the system. 

\begin{figure}[htb]  
   \centering
   \includegraphics[width=8cm]{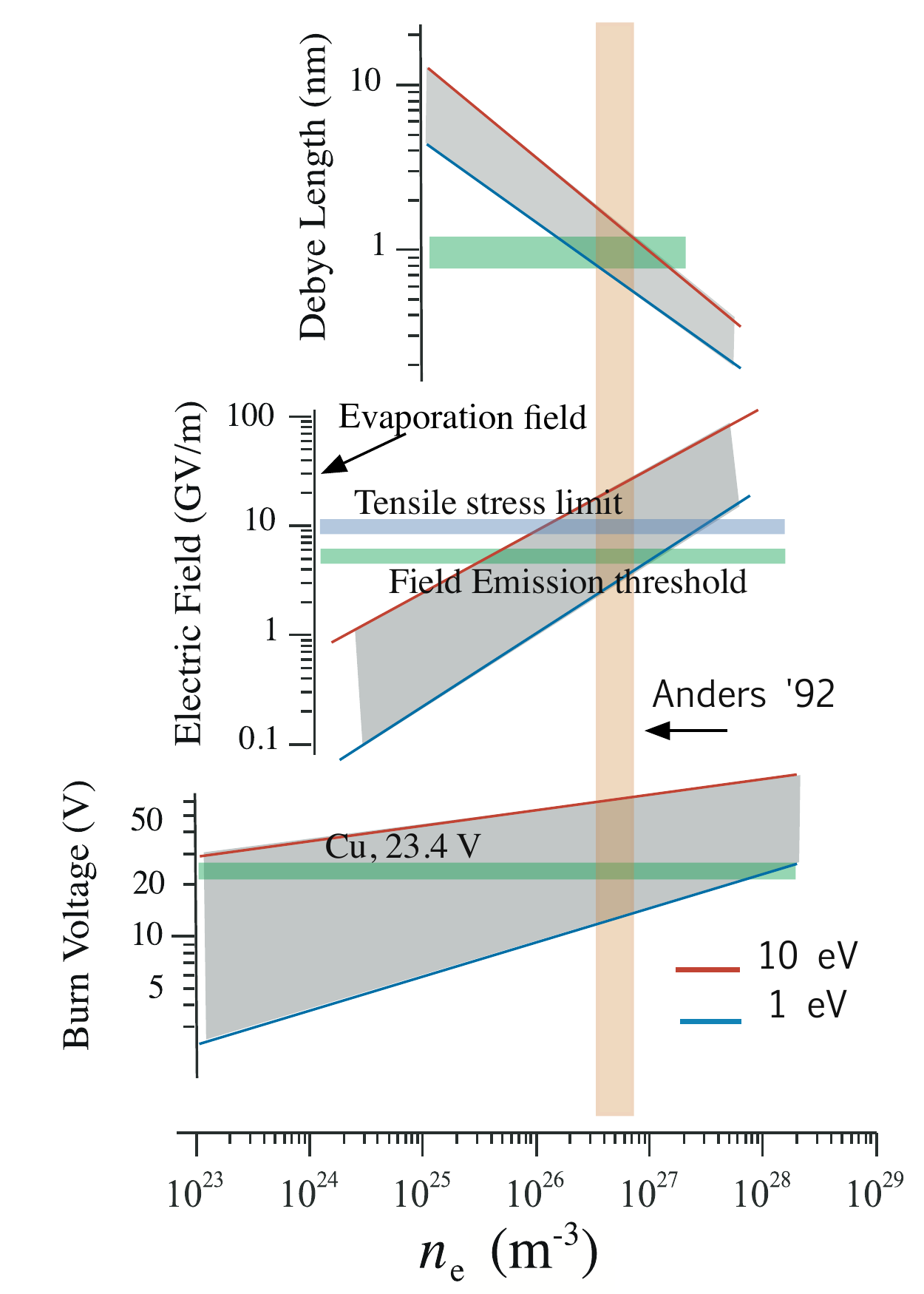}    
   \caption{The interaction between measured properties, (surface field and burn voltage) and electron temperature $T_e$  and density,$n_e$ from sheath calculations.  \cite{morozov,a92}}
\end{figure}    

We also argue that the combination of plasma pressure, Maxwell stresses and surface tension would produce a turbulent liquid surface \cite{IN}.  This turbulence could be a source of instability for the plasma, leading to oscillations, local quenching and plasma motion, with the whole surface under the arc plasma would be emitting dense field emission currents.  

\subsection{Surface Damage}          

We assume that the dominant mechanisms in surface damage are surface tension, local thermal gradients, differential cooling, solidification and contraction at locations of arc and particulate damage.

Unipolar arcs leave a variety of characteristic damage structures on materials \cite{Robson,Schwirzke91}.  The primary mechanisms by which the plasma affects the (presumably molten) metallic surface are plasma pressure caused by ions leaving the plasma and hitting the surface, electrostatic Maxwell stresses pulling on the surface, and surface tension, which tends to locally flatten the surface.  The interplay between these mechanisms is geometry dependent. Plasma pressure tends to be locally constant, however  Maxwell stresses are highly dependent on the local radius of the surface, since the force is dependent on E$^2$, and E near an equipotential is dependent on the local radius.  Surface tension, dependent on the linear dimensions involved, becomes more dominant at small dimensions, unlike pressures, which are dependent on the areas involved.  In general, plasma pressure is similar to hydrostatic pressure, which does not affect the liquid surface geometry, Maxwell stresses pull on convex shapes, becoming stronger as the tips of the surface become sharper.   Surface tension  tries to smooth surfaces.  These effects produce a turbulent surface, perhaps with small areas pulled out from the primary surface.

When the plasma terminates, this turbulent surface relaxes due to surface tension and is governed by capillary waves, locally smoothing the surface \cite{LL}.  At some point the liquid metal will freeze, and the surface will continue to cool, generating stress due to local thermal gradients and differential contraction. This stress is relieved by surface fractures, and  many examples of cracks produced at the center of arc damage sites are seen \cite{IN}.  The cracks produced are seen in SEM images both with and without magnetic fields, and can have the field enhancements required by breakdown calculations, as seen in Fig. 5.  We argue asperities are produced during damage, and there is no need to assume that they grow during subsequent operation of the system.  This growth, which would be detectable in dark current intensities and radiation levels, is not seen \cite{PR1,IN}. 

The production of high field enhancements is a requirement of damage models, in order to predict realistic arcing behavior and conditioning.  These cracks are not the only possible mechanism producing high field enhancements, however.   Systems at higher frequencies that do not see convoluted surfaces, craters or other obvious asperities may not see these crack junctions, but they should be sensitive to particulates which could be deposited and rapidly cooled, leaving small, sharp points (high $\beta$s on an otherwise flat surface) see Fig. 22 of \cite{PR1}, and Fig. 14 of \cite{spitfest}, and Refs. \cite{feynman,ro}. These particulates should be present in vary large numbers, but might require very high magnifications to identify them.

\begin{figure}[htb]  
   \centering
   \includegraphics[width=8cm]{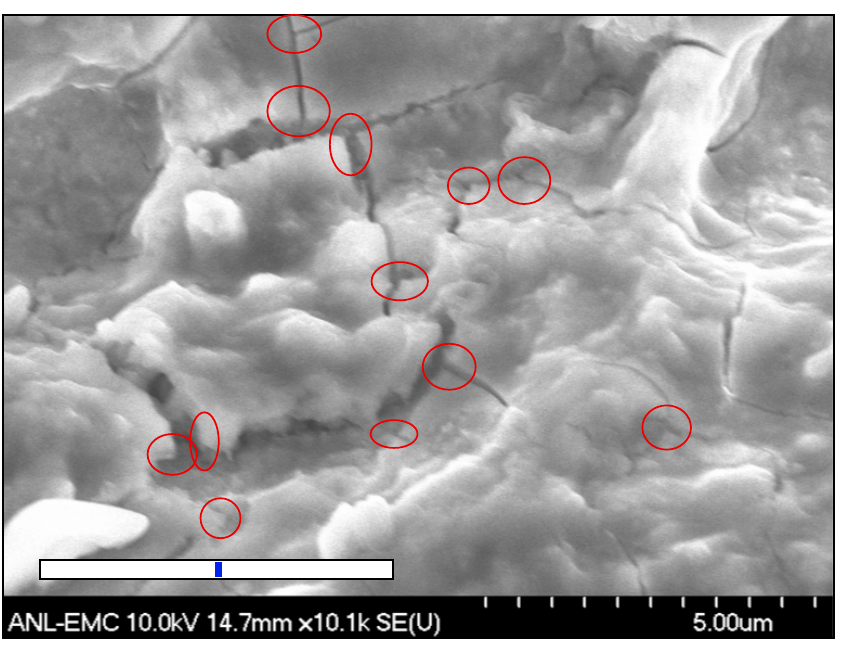}
   \caption{Many cracks visible in SEM images of the center of an arc damage spot at a magnification of 10,100x.  Crack junctions where high field enhancements are expected, are noted. The widths of the cracks are caused by the thermal contraction of the material, $\Delta x \sim x \alpha \Delta T \sim$ 2\% of the initial section, as it cools after solidifying, where $\alpha$ is the coefficient of thermal expansion and $\Delta T$ is the change in temperature. The overall diameter of the damage is $\sim$ 500 $\mu$m, which explains the wider cracks.  The blue spot is 2\% of length the white line for comparison.  The sharp corners would have high local fields, likely sources of field emission   \cite{IN,feynman}. }
\end{figure}

\section{Discussion}            

There are a number of extensions and issues that should be described by any useful model of breakdown.  This field has continued to be very active since J\"{u}ttner's summary was published in 2001, however we would continue to argue that the field is not converging into a simple model of arcing that can explain behavior in many environments. 

\subsection{Diffusion vs. Electromigration}

Unlike diffusion, which is extensively modeled, electromigration has not been studied in connection with breakdown, despite some similarities in the atomic motions required.   Electromigration is described by Black's equation for the Mean Time To Failure in electronic systems, (MTTF), \cite{Black,D,DD,Antoine}
\[ BDR \sim MTTF^{-1} = A  j^2  e^{-Q/kT}\]
 where $A$ is a constant, $j$ is the current density, $Q$ is the activation energy for moving atoms from one place to another, $k$ is Boltzmann's constant and $T$ is the absolute temperature.  Both electromigration and diffusion contain the Arrhenius term, essentially describing activation energy of the same atomic motions, however with electromigration the motions are driven and not random.  The activation energy for atomic motion on the surface is 0.8 V for copper which is lower than other atomic motions, see Table 2.1 in \cite{DD}. Electromigration is the primary cause of failure in electronic components, where it easily dominates diffusion at current densities greater than $\sim 10^{10}$A/m$^2$.  The nonlinear features of electromigration are described in detail in texts on failure modes of microelectronics \cite{D,DD}.
 
\subsection{The Conditioned Surface}          
   
The problem of conditioning illustrates some of the experimental difficulties involved with understanding arcs, with many variables and limited experimental access.  During conditioning, the surface is covered with asperities, and the ones with higher local fields are expected to break down, leaving the maximum local field somewhat lower, permitting an incremental increase in average field.  

Although sufficient data is available to provide a general picture, it comes from many sources and cannot be combined with precision without modeling.  For example, the spectrum of enhancement factors has been measured below the breakdown threshold in a variety of environments \cite{J,hassanein}, field emission from conditioned cavities has been imaged with good resolution \cite{PR1}, and the field dependence of breakdown rate, has been measured \cite{ro}.  However, these measurements have been made on different structures, at different frequencies, in programs with different goals.  Detailed modeling or experimental study of mechanisms is has not been done.
   
If we assume that the breakdown rate is the convolution of the breakdown threshold $t(E_{local})$ and the density distribution of the local field emitters $n(E_{local})$, where the product $\beta E = E_{local}$ is equal to the local field on the field emitters, then the breakdown rate is equal to, 
\[BDR \sim \int n(E_{local})t(E_{local}) dE_{local}.  \]
Although we are unable to measure the function $t(E_{local}$ near the breakdown threshold, it is clear that both $n(E_{local})$ and $t(E_{local})$ must be sharp to produce a dependence like $BDR \sim E^{30}$ \cite{ro}.  

Fig. 6 sketches the spectrum of enhancement factors $n(\beta)$ below the breakdown threshold due to arc damage from previous events.   The emitter density below the breakdown threshold has been measured in unconditioned systems in a number of experiments, giving $n(\beta) \sim e^{-\beta/40}$,  \cite{hassanein} and Fig. 12-14 in \cite{J}.   The dependence of the breakdown rate on field has also been measured above the threshold,  and $BDR \sim E^{30}$, implies that $n(E) \sim E^{-m} $ near the breakdown threshold, as shown in the figure.  The exponent $m$ must be constrained by the relation $m <  n$, otherwise the field emission of asperities far below threshold would overwhelm the asperities near threshold. 

\begin{figure}[htb]  
   \centering
   \includegraphics[width=9cm]{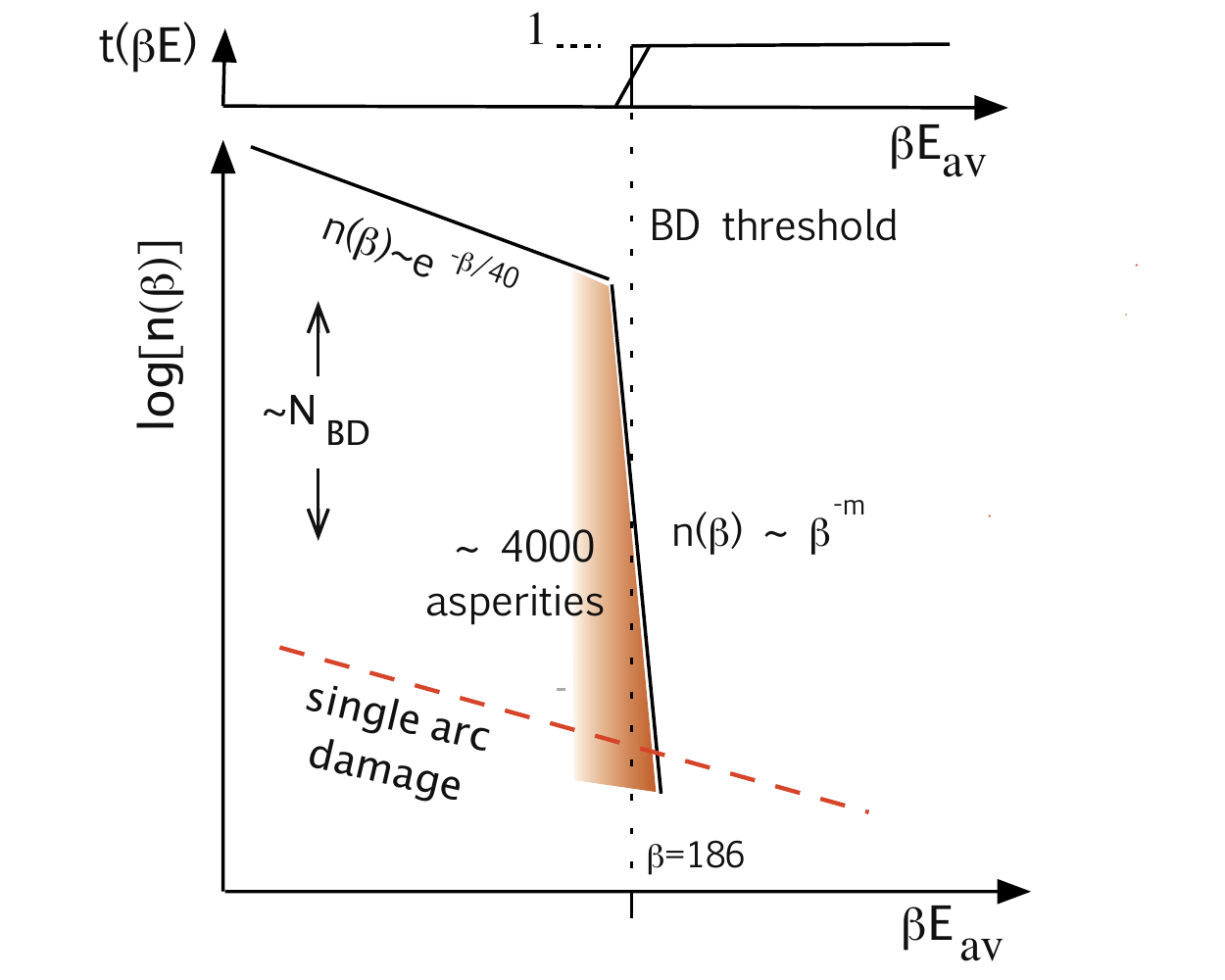}
   \caption{The spectrum, of asperities, $n(\beta )$, with the threshold for breakdown $t(\beta E)$ for a conditioned system \cite{PR1}.  The spectrum of asperities on the arcing surface evolves during conditioning due to creation of asperities with higher $\beta$s, due to single arcs.   These must eventually be burned off by arcing.  This is done at lower voltages to minimize generation of high $\beta$ asperities. Not to scale.}  
\end{figure}

\begin{figure}[htb]  
   \centering
   \includegraphics[width=9cm]{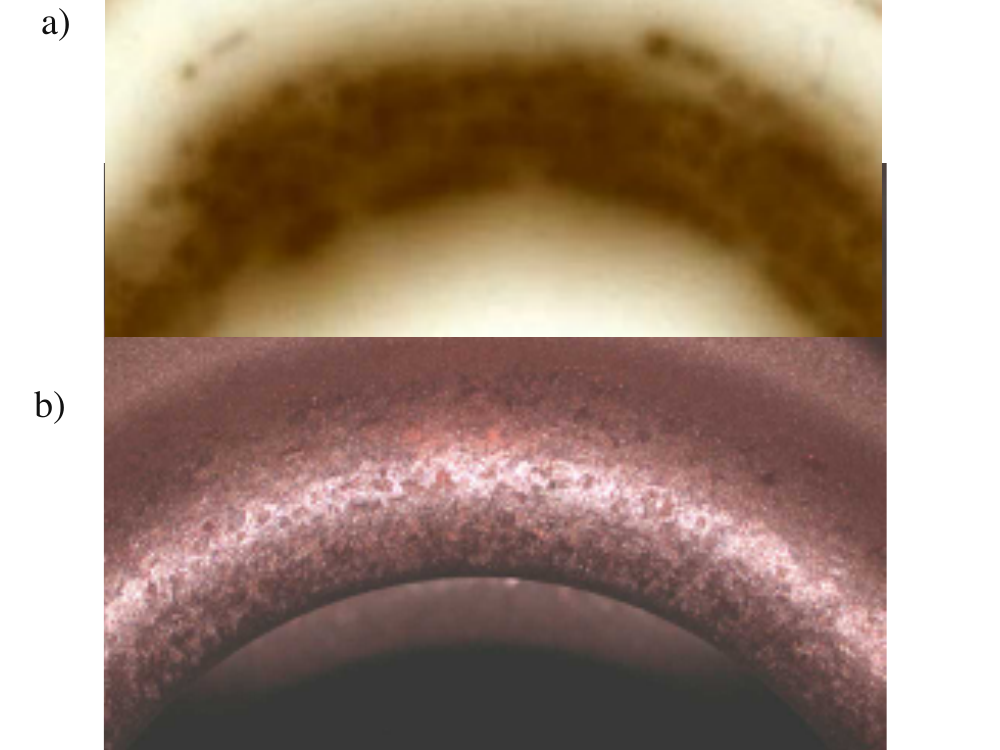}
   \caption{   The active surface in an rf cavity, showing: a) an image of field emitted beamlets from an 8 cm diameter iris in a fully conditioned cavity in a 3 T magnetic field, showing that the surface of the irises is covered in high field enhancement asperities, and the active sites are at essentially equal local field enhancements, and b) an iris in the 6 cell cavity showing arc damage spots about 500 $\mu$m in in diameter.  The emitters seen in Fig 5 are seen in the center of these spots.  Experimental details are presented in \cite{PR1,hassanein},  }
\end{figure}

The colinear magnetic field available in our experiments confined the field emitted electrons near the magnetic  field lines permitting images of the field emitting surface on glass plates and photographic film.
Fig 7 shows a) an image of field emission from the iris of a conditioned cavity showing single damage spots with resolution of $\sim$ 1 mm with a uniform distribution of asperities and, b) a photograph of a conditioned iris in this cavity.   In this large, well conditioned cavity with six irises, with a total field emitting area of approximately 0.054 m$^2$, we estimate a total of about 4000 active damage spots near the breakdown limit  and a density of these spots is roughly 70,000 m$^{-2}$, as seen in Fig. 7a.   This image implies that in a conditioned system, all active emitters have almost identical field enhancements and field emitted currents from single spots would be roughly $1/N_{BD}\sim10^{-3}$, or less, of the total field emitted current, as shown in Fig. 2.

\subsection{Space charge and trigger polarity}

The active areas of emitters should not be smooth, and fluctuations in the local radius should have comparable effects in the local field across the emitter\cite{feynman}.  If the field varied by a factor of three, Fig. 2 shows that current densities could vary by $\sim 10^6 - 10^8$, complicating precise calculations.

The effects of space charge and the angular distribution of emitted electrons from uneven surfaces have been described for many years \cite{dyke}, and frequently modeled \cite{IN}.  When field emitters produce high currents, the electron density near the surface may rise high enough to locally load the electric field to the point where the surface field is significantly reduced,  This field reduction has the effect of limiting both the current density and the surface field available with a negatively charged surface, producing both a space charge limit and a surface field limit, repressing field evaporation and encouraging Joule heating as a failure mechanism.   The field above positively charged surfaces thus could be considerably higher than negatively charged ones, particularly with rough or pointed emitters, see Fig. 13 of \cite{dyke}, with the Maxwell tensile stresses being proportional to the field squared. This argument implies that the trigger of rf breakdown events might most likely be anodic, rather than cathodic so that the breakdown trigger is entirely electrostatic, rather than driven by heating of the surface, which is more likely in DC systems.  This is consistent with the rapid breakdown seen in experiments \cite{spitfest}, and Atom Probe Tomography data, see Fig 5.15 of \cite{F}.

\subsection{Breakdown rates}
Data from breakdown studies show that breakdown rates are not constant but the events come in clusters \cite{spitfest},.  We have shown above that the spectrum of asperities is constrained by the fact that the density of emitters below the breakdown threshold cannot rise faster than the field emission falls, as the field enhancement is reduced, otherwise the field emission would dominate at low enhancements, which is not seen in Fig 7.  This requires that the the breakdown rate rising as $E^{30}$, \cite{spitfest}, is likely due to more than one factor.  We assume that the field dependence of clusters of events must contribute so that,
\[ BDR \sim ( BDR_{of\ cluster} )(BDR_{within\ cluster} ). \]
These event rates are suppressed during conditioning by lowering the applied field, but without lower fields they would proceed at an enhanced rate, producing damage and particulates causing breakdowns in nearby cavities.  This argument implies that the field dependence within clusters of events should be $E^s$, with $s>16$.

\subsection{Emitter Dimensions}
The dimensions of structures emitting dark currents during normal operation can be determined from measurements of surface damage \cite{PR1}.  The damage spots seen in data were all roughly 500 $\mu$m in diameter, where the total field emitting area of the spot is on the order of $2.5 \times 10^{-13}$ m$^2$, from Fig 2.  Assuming 100 - 1000 rectangular corners functioning as emitters, this implies that the individual emitting corners have effective areas of $2.5 - 25 \times 10^{-16}$ m$^2$, or diameters on the order of 16 - 50 nm.   Since the overall dimensions of these structures is on the order of 1 $\mu$m, comparable those seen in \cite{H,dyke}, we expect that these dimensions may be typical for other systems.  Note that when the arc is present the surface fields are large enough to potentially make the whole are area a field emitter.

\section{open questions}            

The model presented here is quite different fro the conventional wisdom.  Much of the data described here has been produced in the same experimental system and one rf frequency.   This model provides an opportunity to reexamine the modeling and experimental details of vacuum arcs from a different and more general perspective.  this model implies that smaller surface structures, a wider range of rf frequencies, and different materials and experimental conditions could all produce useful information.  In addition, it should be useful to look at the similarities between vacuum breakdown, APT surface failure and high current density systems.  

The shape of the spectrum of field enhancements described in this paper has been obtained from a variety of environments, however there should be relevant data in conditioning histories of many cavities  which could be used to verify and improve the work presented here.  More measurements and more modeling of basic parameters such as $T_e, T_i, n_e, n_i$ should be produced.  How do plasma pressure, Maxwell stress and  surface tension determine the time evolution of the surface during and after arcing?

\section{Conclusions}                 

Field gradient limits are central to the design and costing of a number of $\sim$10 B\$ projects and, 120 years after vacuum arcs were identified, these limits are still not well understood.  This paper highlights a number of details not discussed in our many earlier publications (or other work).  Although the expected "unicorn horn" shaped field emitters have not been seen for decades, our data shows surfaces exposed to high electric fields are densely and uniformly covered with field emitters, which seem to be inconspicuous ($\mu$m scale) surface crack junctions.   We show how further study of these issues can help to constrain and advance modeling of vacuum arcs.

\section{Acknowledgements}             

We would like to thank the staff of Fermilab and the management of the Muon Accelerator Program who produced the facility where the experimental work we refer to was done.  This work is based on experiments with 805 MHz cavities sponsored by the US/DOE, and directed at the problem of cooling diffuse beams of muons from accelerator targets.   We would also like to thank  the Center for Nanoscale Materials at Argonne National Lab. including a Hitachi SEM, an Office of Energy, Office of Science, Office of Basic Energy Sciences user facility, under Contract No DE-AC02-06CH11357.  Z.I. is supported in part by the World Science Stars grant by Nazarbayev University.

\end{document}